\documentclass[aps,prx,reprint,amsmath,amssymb,floatfix]{revtex4-2}
\usepackage{microtype}
\usepackage{graphicx}
\usepackage{float}
\usepackage{xcolor}
\usepackage{tabularray}
\UseTblrLibrary{diagbox}
\usepackage{bm}
\usepackage[hidelinks,breaklinks,colorlinks,linkcolor=purple,citecolor=purple,urlcolor=purple]{hyperref}
\usepackage[capitalise,noabbrev]{cleveref}
\usepackage{tikz}
\usetikzlibrary{quantikz2}
\usepackage[sort&compress]{natbib}
\DeclareMathOperator{\diag}{diag}
\DeclareMathOperator{\Pauli}{Pauli}
\DeclareMathOperator{\Orbit}{Orbit}
\renewcommand{\epsilon}{\varepsilon}
\begin{document}
\frenchspacing

\title{Improving error suppression with noise-aware decoding}
\author{Evan T. Hockings}
\email{evan.hockings@sydney.edu.au}
\author{Andrew C. Doherty}
\author{Robin Harper}
\affiliation{ARC Centre of Excellence for Engineered Quantum Systems,\\
School of Physics, The University of Sydney, Sydney, NSW 2006, Australia}
\date{April 1, 2025}

\begin{abstract}
We demonstrate that the performance of quantum error correction can be improved with noise-aware decoders that are calibrated to the likelihood of physical error configurations in a device.
We show that noise-aware decoding increases the error suppression factor of the surface code, yielding reductions in the logical error rate that increase exponentially with the code distance.
Our calibration protocol involves circuit-level Pauli noise characterisation experiments with averaged circuit eigenvalue sampling.
This enables decoder calibration at the scales required for fault-tolerant quantum computation and near-optimal decoding when compared to the true noise model.
Our results indicate that these noise characterisation experiments could be performed and processed in seconds for superconducting quantum computers.
This establishes the practicality and utility of noise-aware decoding for quantum error correction at scale.
\end{abstract}

\maketitle

\textbf{Introduction---}The need for real-time decoding of error syndrome data in fault-tolerant quantum computers will limit decoders and hence the performance of quantum error correction.
Decoders attempt to infer underlying physical errors from error syndrome data to determine a correction operation that preserves the logical information.
They must do so as quickly as the error syndrome data accumulate, or else the backlog grows and slows computation exponentially~\citep{terhal_quantum_2015}.
Consequently, fault-tolerant quantum computation will require fast decoders, such as~\citep{higgott_improved_2023, higgott_sparse_2025, fowler_minimum_2015, wu_fusion_2023}, capable of real-time operation~\citep{barber_realtime_2025, caune_demonstrating_2024}, even if they are less accurate than alternatives~\citep{bravyi_efficient_2014, bausch_learning_2024}.

Mechanisms for improving decoder performance without increasing computational cost are therefore an important factor in reducing fault tolerance overheads.
One such mechanism is noise-aware decoding, which improves performance by calibrating decoders with estimates of the noise in the quantum device.
Calibrating the decoder prior on the likelihood of error configurations allows more accurate inference of physical errors from error syndrome data~\citep{tuckett_faulttolerant_2020, chen_calibrated_2022, sundaresan_demonstrating_2023, tiurev_correcting_2023, higgott_improved_2023}.
Decoder priors can also be learned online from error syndrome data~\citep{spitz_adaptive_2018, wagner_pauli_2022, wang_dgr_2024, remm_experimentally_2025}, or directly optimised to minimise the logical error rate of the code~\citep{sivak_optimization_2024}.

Averaged circuit eigenvalue sampling (ACES)~\citep{flammia_averaged_2022} is a scalable method for characterising the Pauli noise associated with the operation of all gates in the syndrome extraction circuits of large quantum error correcting codes~\citep{hockings_scalable_2025}.
Recently, there has been rapid experimental progress in quantum error correction~\citep{acharya_suppressing_2023, sivak_realtime_2023, bluvstein_logical_2024, dasilva_demonstration_2024, acharya_quantum_2024, caune_demonstrating_2024, reichardt_logical_2024, eickbusch_demonstrating_2024, lacroix_scaling_2024}, generating an immediate need for practical and scalable methods for calibrating the decoder prior.
We use ACES estimates of circuit-level Pauli noise to calibrate a fast correlated matching decoder~\citep{higgott_sparse_2025}.
These estimates can also be used to generate simulated data for training more accurate decoders, such as in~\citep{bausch_learning_2024}, to verify appropriate device calibration, and to inform the tailoring of codes and decoders to noise biases~\citep{tuckett_ultrahigh_2018, tuckett_tailoring_2019, bonillaataides_xzzx_2021, miguel_cellular_2023, dua_clifforddeformed_2024}.

In this letter, we establish the practicality of decoder calibration by demonstrating that ACES enables near-optimal noise-aware decoding when compared to the true noise model.
We show through circuit-level numerical simulations that noise-aware decoding increases the error suppression factor of the surface code~\citep{bravyi_quantum_1998, dennis_topological_2002, kitaev_faulttolerant_2003, fowler_surface_2012}, yielding reductions in the logical error rate that increase exponentially with the code distance.
We find that decoder calibration substantially reduces logical error rates and qubit overheads at the scales necessary for fault-tolerant quantum computation.
Our results indicate that the requisite noise characterisation experiments could be performed and processed in seconds for superconducting quantum computers.
This establishes the practicality and utility of noise-aware decoding for quantum error correction at scale.

\textbf{Introducing ACES---}We now introduce the fundamental ideas underlying averaged circuit eigenvalue sampling (ACES)~\citep{flammia_averaged_2022}.
ACES enables scalable characterisation of the Pauli noise associated with implementing the syndrome extraction circuits of topological quantum codes~\citep{hockings_scalable_2025}.
For a detailed treatment of the ACES protocol used here, refer to~\citep{hockings_scalable_2025}.

Quantum devices typically implement syndrome extraction circuits by performing a sequence of layers of gates.
A \emph{gate} is a primitive operation on the quantum device, typically acting on one or two qubits, and a \emph{layer} is a set of gates that are implemented simultaneously and act on disjoint sets of qubits.
Syndrome extraction circuits are typically implemented with gates drawn from the Clifford group, with some examples being the controlled-\(X\), Hadamard, and phase gates~\citep{aaronson_improved_2004}.
A defining property of the Clifford group is that its elements map Pauli operators to Pauli operators.
Clifford circuits, also known as stabiliser circuits, can be efficiently simulated~\citep{gottesman_stabilizer_1997, aaronson_improved_2004}.
We model noise with Pauli channels
\begin{equation}\label{eq:pauli-channel}
    \mathcal{E}(\rho)=\sum_{\bm{a}\in\mathbf{P}^n}{p_{\bm{a}}P_{\bm{a}}\rho P_{\bm{a}}},
\end{equation}
which act on quantum states by applying a probabilistic mixture of Pauli operators characterised by the \emph{Pauli error probabilities} \(p_{\bm{a}}\).
We represent quantum noise in this way because the technique of Pauli frame randomisation tailors arbitrary noise channels into \emph{Pauli channels}~\citep{knill_quantum_2005, ware_experimental_2021}.

ACES characterises Pauli noise associated with Pauli frame randomised Clifford circuits.
It focuses on a \emph{circuit-level Pauli noise model} by estimating the Pauli error probabilities of the noise channel associated with each gate in the circuit operated in the specific context of its layer.
ACES does this by measuring Pauli observables for circuits arranged from the same layers as the Clifford circuit of interest.
The \emph{circuit eigenvalue} \(\Lambda_\mu\) is precisely the expectation value of a Pauli observable for a rearranged circuit, where \(\mu\) indexes both the Pauli observable and circuit rearrangement.
Classically simulating the action of the circuit on the relevant input states allows us to decompose the circuit eigenvalue into a product of \emph{gate eigenvalues} \(\lambda_\nu\), where \(\nu\) indexes both the gate and the state of the observable before it is acted upon by the gate.
These are, in fact, the eigenvalues of the Pauli noise channel associated with the gate~\citep{flammia_efficient_2020}.
These gate eigenvalues are related to the Pauli error probabilities by an invertible linear transform called the Walsh-Hadamard transform.

Performing stabiliser circuit simulations allows us to turn this into a linear regression problem by creating a \emph{design matrix}.
The design matrix entries \(A_{\mu\nu}\) are the powers of the gate eigenvalue \(\lambda_\nu\) in the decomposition of the circuit eigenvalue \(\Lambda_\mu\).
Taking the logarithm, we obtain the linear system of equations
\begin{equation}\label{eq:design-matrix}
    \log{\Lambda_\mu}=\sum_{\nu}{A_{\mu\nu}\log{\lambda_\nu}}.
\end{equation}
ACES thus reduces the problem of Pauli noise characterisation to estimating a set of circuit eigenvalues such that this design matrix has full rank.
Then we perform linear least squares to obtain estimates of the gate eigenvalues and hence the Pauli error probabilities of each of the gates in the circuit of interest.
We can optimise the design of ACES noise characterisation experiments to improve the sample efficiency of their noise estimates, as described in~\citep{hockings_scalable_2025}.
\Cref{apdx:relative-aces} outlines a number of improvements to the ACES protocol of~\citep{hockings_scalable_2025}.

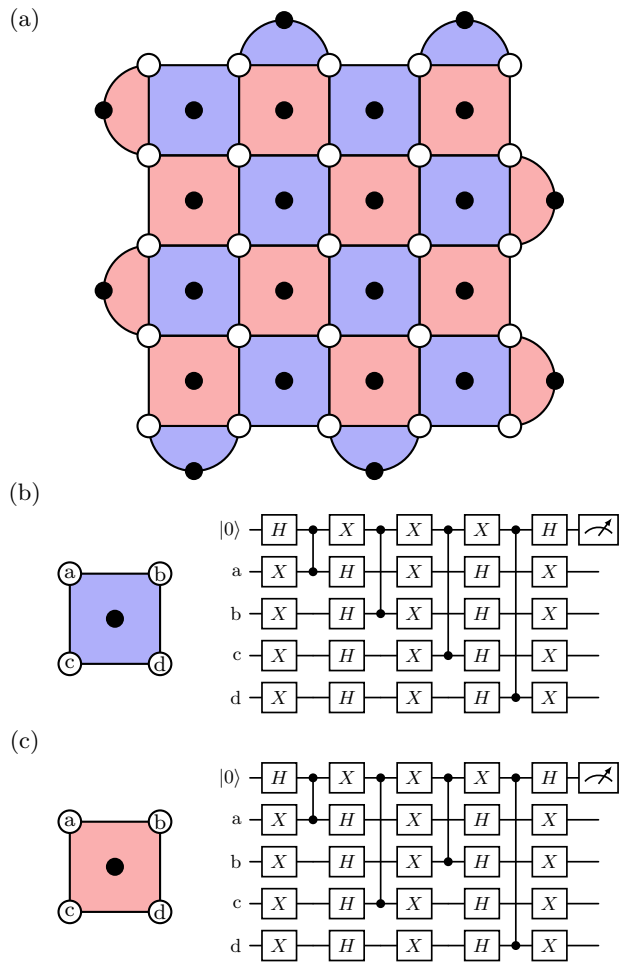
\begin{figure}
    \raggedright
    \definecolor{lightred}{RGB}{250, 175, 175}
    \definecolor{lightblue}{RGB}{175, 175, 250}
    \begin{tikzpicture}[scale=0.6]
        \newcommand \x {1.75};
        \newcommand \y {11.5};
        \node at (0,\y+10) {(a)};
        \fill[lightblue] (\x+2,\y+1) circle (1);
        \fill[lightblue] (\x+6,\y+1) circle (1);
        \draw[thick] (\x+2,\y+1) circle (1);
        \draw[thick] (\x+6,\y+1) circle (1);
        \fill[black] (\x+2,\y+0) circle (0.2);
        \fill[black] (\x+6,\y+0) circle (0.2);
        \fill[lightblue] (\x+4,\y+9) circle (1);
        \fill[lightblue] (\x+8,\y+9) circle (1);
        \draw[thick] (\x+4,\y+9) circle (1);
        \draw[thick] (\x+8,\y+9) circle (1);
        \fill[black] (\x+4,\y+10) circle (0.2);
        \fill[black] (\x+8,\y+10) circle (0.2);
        \fill[lightred] (\x+1,\y+4) circle (1);
        \fill[lightred] (\x+1,\y+8) circle (1);
        \draw[thick] (\x+1,\y+4) circle (1);
        \draw[thick] (\x+1,\y+8) circle (1);
        \fill[black] (\x+0,\y+4) circle (0.2);
        \fill[black] (\x+0,\y+8) circle (0.2);
        \fill[lightred] (\x+9,\y+2) circle (1);
        \fill[lightred] (\x+9,\y+6) circle (1);
        \draw[thick] (\x+9,\y+2) circle (1);
        \draw[thick] (\x+9,\y+6) circle (1);
        \fill[black] (\x+10,\y+2) circle (0.2);
        \fill[black] (\x+10,\y+6) circle (0.2);
        \foreach \i in {1,...,4} {
            \foreach \j in {1,...,4} {
                \ifnum\ifodd\numexpr\i+\j\relax 0\else 1\fi=1
                    \fill[lightred] (\x+\i*2-1,\y+\j*2-1) rectangle (\x+\i*2+1,\y+\j*2+1);
                \else
                    \fill[lightblue] (\x+\i*2-1,\y+\j*2-1) rectangle (\x+\i*2+1,\y+\j*2+1);
                \fi
            }
        }
        \foreach \i in {1,...,4} {
            \foreach \j in {1,...,4} {
                \draw[thick] (\x+\i*2-1,\y+\j*2-1) rectangle (\x+\i*2+1,\y+\j*2+1);
            }
        }
        \foreach \i in {0,...,4} {
            \foreach \j in {0,...,4} {
                \fill[white] (\x+\i*2+1,\y+\j*2+1) circle (0.25);
                \draw[thick] (\x+\i*2+1,\y+\j*2+1) circle (0.25);
            }
        }
        \foreach \i in {1,...,4} {
            \foreach \j in {1,...,4} {
                \fill[black] (\x+\i*2,\y+\j*2) circle (0.2);
            }
        }
        \newcommand \z {1};
        \node at (0,11) {(b)};
        \fill[lightblue] (\z,5.5+1.725) rectangle (\z+2,5.5+1.725+2);
        \draw[thick] (\z,5.5+1.725) rectangle (\z+2,5.5+1.725+2);
        \foreach \i in {0,1} {
            \foreach \j in {0,1} {
                \fill[white] (\z+\i*2,5.5+1.725+\j*2) circle (0.25);
                \draw[thick] (\z+\i*2,5.5+1.725+\j*2) circle (0.25);
            }
        }
        \fill[black] (\z+1,5.5+1.725+1) circle (0.2);
        \node at (\z,5.5+1.725+2) {\footnotesize{a}};
        \node at (\z+2,5.5+1.725+2) {\footnotesize{b}};
        \node at (\z,5.5+1.725) {\footnotesize{c}};
        \node at (\z+2,5.5+1.725) {\footnotesize{d}};
        \node[scale=0.8,anchor=north west] at (4,11) {
        \begin{quantikz}[column sep=0.2cm,row sep=0.2cm]
            \lstick{\(|0\rangle\)} & \gate{H} & \ctrl{1} & \gate{X} & \ctrl{2} & \gate{X}
            & \ctrl{3} & \gate{X} & \ctrl{4} & \gate{H} & \meter{}\\
            \lstick{a} & \gate{X} & \control{} & \gate{H} & & \gate{X} & & \gate{H} & & \gate{X} &\\
            \lstick{b} & \gate{X} & & \gate{H} & \control{} & \gate{X} & & \gate{H} & & \gate{X} &\\
            \lstick{c} & \gate{X} & & \gate{H} & & \gate{X} & \control{} & \gate{H} & & \gate{X} &\\
            \lstick{d} & \gate{X} & & \gate{H} & & \gate{X} & & \gate{H} & \control{} & \gate{X} &
        \end{quantikz}
        };
        \node at (0,5.5) {(c)};
        \fill[lightred] (\z,1.725) rectangle (\z+2,1.725+2);
        \draw[thick] (\z,1.725) rectangle (\z+2,1.725+2);
        \foreach \i in {0,1} {
            \foreach \j in {0,1} {
                \fill[white] (\z+\i*2,1.725+\j*2) circle (0.25);
                \draw[thick] (\z+\i*2,1.725+\j*2) circle (0.25);
            }
        }
        \fill[black] (\z+1,1.725+1) circle (0.2);
        \node at (\z,1.725+2) {\footnotesize{a}};
        \node at (\z+2,1.725+2) {\footnotesize{b}};
        \node at (\z,1.725) {\footnotesize{c}};
        \node at (\z+2,1.725) {\footnotesize{d}};
        \node[scale=0.8,anchor=north west] at (4,5.5) {
        \begin{quantikz}[column sep=0.2cm,row sep=0.2cm]
            \lstick{\(|0\rangle\)} & \gate{H} & \ctrl{1} & \gate{X} & \ctrl{3} & \gate{X} & \ctrl{2} & \gate{X} & \ctrl{4} & \gate{H} & \meter{}\\
            \lstick{a} & \gate{X} & \control{} & \gate{H} & & \gate{X} & & \gate{H} & & \gate{X} &\\
            \lstick{b} & \gate{X} & & \gate{H} & & \gate{X} & \control{} & \gate{H} & & \gate{X} &\\
            \lstick{c} & \gate{X} & & \gate{H} & \control{} & \gate{X} & & \gate{H} & & \gate{X} &\\
            \lstick{d} & \gate{X} & & \gate{H} & & \gate{X} & & \gate{H} & \control{} & \gate{X} &
        \end{quantikz}
        };
    \end{tikzpicture}
    \caption{(a) Diagram of a distance \(d=5\) rotated surface code.
    Data qubits are shown as open circles, and measure qubits are shown as filled circles placed in their corresponding stabiliser plaquettes.
    Syndrome extraction circuit components are shown for (b) blue and (c) red stabiliser plaquettes, respectively.
    These circuits implement the XZZX variant of the surface code~\citep{wen_quantum_2003, bonillaataides_xzzx_2021}, following recent experiments~\citep{acharya_suppressing_2023, acharya_quantum_2024}.}
    \label{fig:rotated-code-circuit}
\end{figure}

\textbf{Quantum codes and decoders---}Quantum error correcting codes typically work by repeatedly performing parity check measurements~\citep{terhal_quantum_2015}, which may or may not mutually commute in the respective cases of stabiliser codes~\citep{gottesman_stabilizer_1997} and subsystem codes~\citep{poulin_stabilizer_2005}.
These parity check measurements are implemented by repeatedly performing a \emph{syndrome extraction circuit}.
A \emph{detector} of such a circuit is a parity of measurement outcomes---a product of parity check measurements---that is deterministic in the absence of noise.
A detector is said to be flipped if its value differs from the deterministic outcome.

We focus on surface codes~\citep{bravyi_quantum_1998, dennis_topological_2002, kitaev_faulttolerant_2003, fowler_surface_2012}, a well-studied family of stabiliser codes that encode a single logical qubit in a \(d\times d\) square array of physical qubits called data qubits.
The parameter \(d\) is the distance of the code.
A distance \(d=5\) surface code is depicted in \cref{fig:rotated-code-circuit} alongside the syndrome extraction circuit we consider in this letter.
Also shown in \cref{fig:rotated-code-circuit} are the \(d^2-1\) measure qubits used by this circuit to measure the Pauli \(Z\) and \(X\) stabilisers of the code.
We operate the surface code in a memory experiment by repeatedly performing its syndrome extraction circuit.
The detectors in a memory experiment are the parities of stabiliser measurement outcomes across consecutive rounds of syndrome extraction.

Stabiliser simulation of the syndrome extraction circuit with a circuit-level Pauli noise model, which ACES can estimate, allows us to efficiently calculate the \emph{decoder prior}.
The decoder prior describes the probability of each possible combination of detector flips caused by Pauli error mechanisms in the circuit.
Then the \emph{syndrome} of an error is the set of detectors flipped by that error.

A decoder infers a correction operation from this syndrome data that aims to preserve the logical information in the code.
Fast decoders avoid the decoding backlog problem~\citep{terhal_quantum_2015}, enabling quantum error correction at scale.
We focus on fast minimum-weight perfect matching (MWPM) decoders~\citep{dennis_topological_2002, fowler_minimum_2015, higgott_sparse_2025, wu_fusion_2023}, which can operate in constant parallel time~\citep{fowler_minimum_2015, wu_fusion_2023}.
MWPM decoders infer the most probable error consistent with the syndrome.
This probability is defined by the decoder prior, and so accurate calibration of this prior is important---not only for MWPM decoders but indeed for all decoders.

A memory experiment examines how well a quantum error correcting code preserves logical information, such as Pauli \(X\) or \(Z\) logical observables.
We will examine surface code memory experiments characterised by the code distance \(d\) and rounds of syndrome extraction \(r\), following the circuits in~\citep{acharya_suppressing_2023, acharya_quantum_2024}.
When the noise associated with implementing the syndrome extraction circuit is sufficiently low, the code is said to be operating below its threshold~\citep{fowler_surface_2012}.
Below threshold, the logical error rate per round of syndrome extraction \(\epsilon\) is well-described by the approximate scaling relation \(\epsilon\propto\Lambda^{-d/2}\).
The \emph{error suppression factor} \(\Lambda\) characterises the error suppression associated with increasing the code distance by \(2\) and depends on the decoder.
We are interested in comparing the error suppression factors when the decoder is calibrated with different priors, as error suppression factor improvements correspond to exponential reductions in the logical error rate.

\textbf{Numerical results---}We now present numerical results demonstrating the utility of calibrating the decoder prior and the practicality of doing so with ACES.
These results were produced on a 2021 M1 Max Macbook Pro with 32 GB of RAM.
We use Stim for stabiliser circuit simulations~\citep{gidney_stim_2021}, PyMatching for a fast correlated matching decoder~\citep{higgott_pymatching_2022, higgott_sparse_2025}, and QuantumACES to design and simulate ACES noise characterisation experiments~\citep{hockings_quantumacesjl_2025}.

We test our methods on a distribution over noise models called log-normal Pauli noise, where the Pauli error probabilities of each gate are independently log-normally distributed, following~\citep{hockings_scalable_2025}.
We use a physically relevant noise parameter regime resembling a recent experiment that operated the surface code below threshold~\citep{acharya_quantum_2024}.
In particular, we choose average single-qubit, two-qubit, measurement and measurement idle, and reset error rates of \(r_1=0.05\%\), \(r_2=0.4\%\), \(r_m=0.8\%\), and \(r_r=0.2\%\), and standard deviations \(\sigma_1=\sigma_2=1/2\) and \(\sigma_m=\sigma_r=1/4\) for the underlying normal distributions.
For data reproducibility, we fix the seed to be \(0\) when drawing only a single random instance from this distribution over noise models.
We will also consider tuned depolarising noise, where component error rates are set to be their average values for log-normal Pauli noise, and optimise our ACES experimental design according to these parameters for a distance-\(3\) surface code syndrome extraction circuit.

First, we examine the performance of surface code memory experiments, averaged over random instances of log-normal Pauli noise, across a range of decoder priors and code distances.
These results were obtained by fitting the exponentially decaying logical error rate, estimated over \(10^5\) shots, over rounds \(r\in\{3,5,9,17,33\}\) in both \(X\) and \(Z\) memory experiments.
We average over both the memory experiment type and \(\{1500,300,100,80,60,50\}\) random instances of log-normal Pauli noise for distances \(\{3,5,7,9,11,13\}\), respectively.
In each instance, decoding was performed with priors derived from the true noise model, tuned depolarising noise, and ACES noise estimates with \(10^6\) and \(10^7\) shots.
Fitting the logical error per round as an exponential in the code distance yields average error suppression factor estimates \(\Lambda_{\mathrm{true}}=1.7360\pm 0.0025\) for the true noise model, \(\Lambda_{\mathrm{dep}}=1.6967\pm 0.0025\) for tuned depolarising noise, and \(\Lambda_{\mathrm{ACES}:10^6}=1.7347\pm 0.0025\) and \(\Lambda_{\mathrm{ACES}:10^7}=1.7358\pm 0.0025\) for ACES noise estimates with \(10^6\) and \(10^7\) shots, respectively.
We find that decoding with ACES noise estimates performs similarly to decoding with the true noise model, even with only \(10^6\) shots, whereas decoding with depolarising noise performs significantly worse.

Importantly, noise-aware decoding increases the error suppression factor, leading to reductions in the logical error rate that increase exponentially with the code distance.
Intuitively, as the code distance increases, tolerable error configurations become larger and more complicated, increasing the freedom in decoding.
Hence we would expect the benefit of calibrating the decoder prior to increase with scale as we see here.
This motivates a focus on highly scalable methods for calibrating decoder priors such as ACES.

\begin{figure}[t]
    \centering
    \includegraphics[width=\columnwidth]{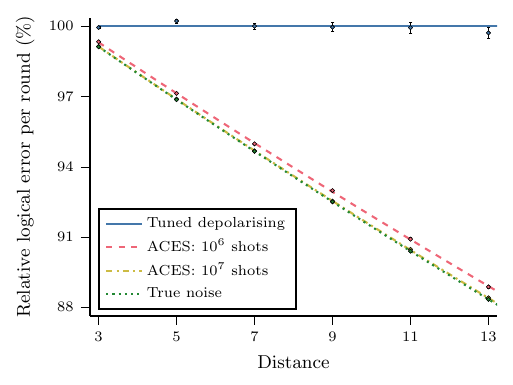}
    \vspace{-2em}
    \caption{Average logical error rate per round of surface code syndrome extraction across priors for the MWPM decoder expressed relative to tuned depolarising noise as a function of the code distance.
    Results are averaged over random instances of log-normal Pauli noise and \(X\) and \(Z\) memory experiments and the MWPM decoder prior is derived from the true noise model, tuned depolarising noise, and ACES noise estimates obtained with \(10^6\) and \(10^7\) shots.
    The data points are fit to an exponential in the code distance, error bars indicate one standard deviation, and all values are expressed with respect to the fit for tuned depolarising noise.
    The performance gain from decoding with ACES noise estimates rather than tuned depolarising noise grows linearly with the code distance, and ACES noise estimates with \(10^7\) shots obtain almost all of the performance gains associated with decoding with the true noise model.}
    \label{fig:aces-ratio}
\end{figure}

\Cref{fig:aces-ratio} displays the relative performance of different decoder priors averaged over random instances of log-normal Pauli noise, expressing the logical error per round for different decoder priors relative to tuned depolarising noise.
The covariance of the logical error rates between decoder priors across each instance of log-normal Pauli noise allows us to precisely estimate error suppression factors relative to the true noise model prior.
Then the data are normalised by the fit for tuned depolarising noise to highlight the performance gains of noise-aware decoding.
These results show that decoding with ACES noise estimates allows near-optimal decoding as compared to the true noise model, whereas decoding with tuned depolarising noise noticeably decreases performance.

\begin{table}[t]
    \centering
    \caption{Decoder performance for distance-\(25\) surface code memory \(X\) and \(Z\) experiments with \(25\) rounds under a random instance of log-normal Pauli noise, dividing \(10^7\) shots evenly between the memory types.
    The MWPM decoder prior is derived from the true noise model, ACES noise estimates with \(10^7\) and \(10^6\) shots, and tuned depolarising noise, ordered by the accuracy and decoding performance of the noise model.
    Diagonal elements count decoding failures for each prior.
    Off-diagonal elements count the number of shots where the decoder for the row succeeded and the decoder for the column failed.
    Decoding with ACES noise estimates achieves similar logical error rates to decoding with the true noise model.}
    \vspace{0.25em}
    \def\arraystretch{1.4}
    \begin{tblr}{column{1}={l,m},column{2-5}={c,m,colsep=3.4pt},cell{1}{2-5}={c,h},hlines,vlines}
        \diagboxthree{Succ.}{\textbf{Fail.}}{Fail.} 
                      & \raisebox{-1pt}{True} & \raisebox{-1pt}{ACES:\(10^7\)} 
                      & \raisebox{-1pt}{ACES:\(10^6\)} & \raisebox{-1pt}{Depolarising} \\
        True          & \textbf{5507} & 227           & 619           & 3005          \\
        ACES:\(10^7\) & 195           & \textbf{5539} & 564           & 2997          \\
        ACES:\(10^6\) & 495           & 472           & \textbf{5631} & 2994          \\
        Depolarising  & 1314          & 1338          & 1427          & \textbf{7198} \\
    \end{tblr}
    \label{tab:decode-confusion}
\end{table}

Next, we test noise-aware decoding at scale by examining surface code memory experiments at the larger distance \(d=25\).
To begin, we simulate a memory experiment for a single random instance of log-normal Pauli noise with \(r=25\) rounds, dividing \(10^7\) shots evenly between \(X\) and \(Z\) memory experiments.
The decoder confusion matrix in \cref{tab:decode-confusion} compares the raw error counts when decoding with the true noise model, tuned depolarising noise, and weighted least squares ACES noise estimates with \(10^6\) and \(10^7\) shots.
Each comparison between decoder priors shows it is more common for decoding with the more accurate noise model to succeed when the less accurate noise model fails, though the converse does occur.

We also fit the logical error per round estimated over rounds \(r\in\{3,5,9,17,33\}\) with \(10^6\) shots for both \(X\) and \(Z\) memory experiments, averaging over the memory experiment type.
We obtain logical error per round estimates \(\epsilon_{25,\mathrm{true}}=(2.39\pm 0.05)\times 10^{-5}\) for the true noise model, \(\epsilon_{25,\mathrm{dep}}=(3.13\pm 0.06)\times 10^{-5}\) for tuned depolarising noise, and \(\epsilon_{25,\mathrm{ACES}:10^6}=(2.42\pm 0.05)\times 10^{-5}\) and \(\epsilon_{25,\mathrm{ACES}:10^7}=(2.40\pm 0.05)\times 10^{-5}\) for ACES noise estimates with \(10^6\) and \(10^7\) shots, respectively.
Now we can compare relative logical errors per round for this single random instance of log-normal Pauli noise to the extrapolated trends of \cref{fig:aces-ratio} which instead describe the average over random instances of log-normal Pauli noise.
We have \(\epsilon_{25,\mathrm{dep}}/\epsilon_{25,\mathrm{true}}=1.31\pm 0.04\) compared to the predicted average value \(1.300\pm 0.004\), \(\epsilon_{25,\mathrm{ACES}:10^6}/\epsilon_{25,\mathrm{true}}=1.01\pm 0.03\) compared to \(1.012\pm 0.001\), and \(\epsilon_{25,\mathrm{ACES}:10^7}/\epsilon_{25,\mathrm{true}}=1.00\pm 0.03\) compared to \(1.001\pm 0.001\).
Hence the improvement from noise-aware decoding for this single random instance of log-normal Pauli noise at a large scale is highly consistent with the extrapolated trends for the average over instances of log-normal Pauli noise at small scales.
We attribute this to a self-averaging effect.

Given this confidence in the extrapolated trends of \cref{fig:aces-ratio}, we can now examine the logical error rate and qubit reductions achievable at the scales necessary for quantum computation.
For example, at distance \(d=63\), decoding with tuned depolarising noise is predicted to obtain a logical error rate of approximately \(\epsilon=10^{-9}\).
We find \(\epsilon_{63,\mathrm{true}}=(5.169\pm 0.009)\times 10^{-10}\) for the true noise model, \(\epsilon_{63,\mathrm{dep}}=(10.037\pm 0.019)\times 10^{-10}\) for tuned depolarising noise, and \(\epsilon_{63,\mathrm{ACES}:10^6}=(5.296\pm 0.009)\times 10^{-10}\) and \(\epsilon_{63,\mathrm{ACES}:10^7}=(5.191\pm 0.009)\times 10^{-10}\) for ACES noise estimates with \(10^6\) and \(10^7\) shots, respectively.
At this distance, noise-aware decoding is predicted to roughly halve the logical error rate.
Indeed, at distance \(d=61\), we predict \(\epsilon_{61,\mathrm{ACES}:10^6}=(9.187\pm 0.015)\times 10^{-10}\).
This represents a simultaneous reduction in the logical error rate and qubit overhead, from \(n=7937\) at \(d=63\) to \(n=7441\) at \(d=61\), a relative qubit reduction of over \(6\%\) and an absolute reduction of \(496\) qubits.

\textbf{Conclusions---}In this letter, we have demonstrated that ACES is practically capable of calibrating decoder priors.
We find that noise-aware decoding increases the error suppression factor of the surface code, exponentially reducing logical error rates compared to decoding with a tuned depolarising noise model.
Calibration with ACES noise estimates enables near-optimal decoding compared to calibration with the true noise model.
Our findings indicate that noise-aware decoding can substantially reduce logical error rates and qubit overheads at the scales necessary for quantum computation.
As the tuned depolarising noise model we use as a comparison is itself informed by accurate estimates of the average error rates, gains may be larger in practice.

Noise estimates obtained from ACES with \(10^6\) shots appear to be sufficiently precise for the purpose of calibrating decoder priors.
In the surface code experiment referenced for our noise model parameter regime~\citep{acharya_quantum_2024}, gate, measurement, and reset times indicate that ACES experiments with the design featured in this letter could collect \(10^6\) shots in about two seconds with an appropriate hardware and software control stack.
In our laptop-based numerical simulations of ACES for a distance-\(25\) surface code syndrome extraction circuit, classical processing of the data with ACES can be performed in under four seconds.
This processing time could be substantially reduced by improved hardware and optimised software.
Therefore, we expect that decoder priors could be calibrated with ACES experiments performed and processed in seconds for superconducting quantum computers.

A natural next step would be implementing these methods in a memory experiment performed on a quantum device.
However, as we find that the key benefit of noise-aware decoding is in improving the error suppression factor of the code, performance improvements may be difficult to realise in small codes.
It would be instructive to more thoroughly examine the calibration of decoder priors with ACES across different error rates and decoders.
It would also be interesting to compare the performance and practicality of calibration methods based on noise characterisation, such as ACES, with those that directly attempt to minimise the logical error rate, for example with reinforcement learning~\citep{sivak_optimization_2024}.
Moreover, ACES noise estimates could be used to inform simulations that generate synthetic data that could then be included in the training data for neural network decoders, such as~\citep{bausch_learning_2024}, which may improve performance.

An important advantage of methods such as ACES, which use estimates of circuit-level noise to calibrate decoder priors, is that they can also diagnose device performance and identify poorly performing gates and measurements.
Decoder priors calibrated in this way could then be updated online using syndrome data~\citep{spitz_adaptive_2018, wagner_pauli_2022, wang_dgr_2024, remm_experimentally_2025}, accounting for device drift.
We hope this work paves the way for a sophisticated yet practical stack of methods for rapid calibration of decoder priors at scale in fault-tolerant quantum computing architectures.

\textbf{Acknowledgements---}We thank Nicholas Fazio for discussions.
This work was supported by the Australian Research Council Centre of Excellence for Engineered Quantum Systems (CE170100009) and the U.S. Army Research Office (W911NF-23-S-0004).

\textbf{Data availability---}The code supporting the findings of this letter is openly available in QuantumACES~\citep{hockings_quantumacesjl_2025}.

\bibliography{aces-decoding}

\clearpage

\appendix

\section{Relative precision ACES}\label{apdx:relative-aces}

In this appendix, we briefly outline a number of modifications we make to the ACES protocol described in~\citep{hockings_scalable_2025}, enabling relative precision noise estimation.
We build upon the notation of~\citep{hockings_scalable_2025} and do not separately introduce it here.
Our modifications follow Pauli noise characterisation theory~\citep{chen_learnability_2023, chen_efficient_2024}, also outlined in cycle error reconstruction~\citep{carignan-dugas_error_2023}.

The key idea is to estimate gate eigenvalues corresponding to the \emph{orbits} of a gate, sets of Paulis mapped to each other by the action of the gate, rather than for all of the Paulis.
Given some Clifford gate \(\mathcal{G}\), let the \emph{\(\mathcal{G}\)-orbit} of a Pauli \(P_{\bm{a}}\) be
\begin{equation}\label{eq:gate-orbit}
    P_{\bm{a}}^{\mathcal{G}}=\{\mathcal{G}^{j}(P_{\bm{a}})\colon j\in\mathbb{N}\},
\end{equation}
where we neglect the sign of the Paulis.
Then the \(\mathcal{G}\)-\emph{orbits} are simply the \(\mathcal{G}\)-orbits of each of the Paulis supported on \(\mathcal{G}\), which we write as
\begin{equation}\label{eq:gate-orbits}
    \Orbit{(\mathcal{G})}=\{P_{\bm{a}}^{\mathcal{G}}\colon\bm{a}\in\Pauli{(\mathcal{G})}\}.
\end{equation}
Note that \(|\!\Orbit{(\mathcal{G})}|\le|\!\Pauli{(\mathcal{G})}|\), which is strict if and only if there exist Paulis on which the gate acts non-trivially up to sign, so that equality is achieved by Pauli gates.

Now consider the circuit eigenvalue corresponding to preparing the Pauli \(P_{\bm{a}}\) and then repeating the gate \(\mathcal{G}\) \(t\) times, which we will write as \(\Lambda_{\mathcal{G}^{(t)},\bm{a}}\).
Now divide \(t\) by the size of the \(\mathcal{G}\)-orbit of \(P_{\bm{a}}\) to obtain integers \(u,v\) such that \(t=|P_{\bm{a}}^{\mathcal{G}}|u+v\) where \(v<|P_{\bm{a}}^{\mathcal{G}}|\).
Then we can write
\begin{equation}\label{eq:orbit-circuit-eigenvalue}
    \Lambda_{\mathcal{G}^{(t)},\bm{a}}=\left(\prod_{\bm{a}\in P_{\bm{a}}^{\mathcal{G}}}{\lambda_{\mathcal{G},\bm{a}}}\right)^{u}\lambda_{G,\mathcal{G}^{0}(\bm{a})}\dots\lambda_{\mathcal{G},\mathcal{G}^{v}(\bm{a})}.
\end{equation}
We see that circuit eigenvalues with this form allow precise estimation of the product of gate eigenvalues in the \(\mathcal{G}\)-orbit of \(P_{\bm{a}}\).
Indeed, this product can be estimated to \emph{relative} precision by repeating the gate a number of times inversely proportional to the noise strength.
By contrast, the individual gate eigenvalues within the orbit can only be estimated to \emph{additive} precision.

Examining the covariance structure of the gate eigenvalue estimators in optimised ACES experimental designs reveals that ACES naturally exhibits this behaviour without modification.
While the estimators for gate eigenvalues within an orbit have large variance, the covariance of these estimators ensures that the average over the orbit has small variance.
This can be revealed by marginalising the gate eigenvalue estimator covariance matrix \(\Sigma\) over the gate orbits and ignoring parameters that can only be estimated to additive precision, such as those corresponding to state preparation and measurement (SPAM) noise.
This yields the relative precision gate eigenvalue estimator covariance matrix \(\Sigma_R\).
Replacing \(\Sigma\) with \(\Sigma_R\) in the expression for the ordinary precision figure of merit \(\mathcal{F}\) in~\citep{hockings_scalable_2025} allows us to calculate the relative precision figure of merit \(\mathcal{F}_R\).

In light of this, we make two changes to how we optimise ACES experimental designs.
We optimise the depth of repeated circuits according to this relative precision figure of merit, whereas we still optimise the shallow circuits according to the ordinary precision figure of merit.
The ordinary and relative precision figures of merit trade off against each other, which we manage by optimising the shot weights according to the product \(\mathcal{F}\mathcal{F}_R\), or equivalently their geometric mean.
This balances the relative increase in one figure of merit with the relative decrease in the other.

However, while ACES estimates the gate eigenvalues, and hence the gate error probabilities, to relative precision, this is not necessarily true after projecting the estimator into the probability simplex in the Euclidean distance.
To preserve the estimator covariance information necessary for relative precision estimation, we instead project in the \emph{Mahalanobis distance}~\citep{gelman_causal_2006}, which weights the distance with the inverse of the gate probability estimator covariance matrix \(\Sigma_{P}^{-1}\).
This enables relative precision estimation by retaining estimator covariance information, and also improves the accuracy of the resulting estimator compared to Euclidean distance projection.

Let us first consider the inverse of the gate eigenvalue estimator covariance matrix \(\Sigma^{-1}\), which for the generalised least squares estimator can be calculated as
\begin{equation}\label{eq:eigenvalue-precision}
    \Sigma^{-1}=\diag{(\bm{\lambda}^{-1})}A^\intercal\Omega^{\prime -1} A\diag{(\bm{\lambda}^{-1})}.
\end{equation}
This calculation preserves the sparse structure of the design matrix \(A\) and the circuit log-eigenvalue estimator covariance matrix \(\Omega^\prime\).
We perform the calculation in \cref{eq:eigenvalue-precision} using the gate eigenvalue estimator vector \(\hat{\bm{\lambda}}\) and the estimator \(\hat{\Omega}^\prime\), whose elements can be calculated directly from measurement outcome data.
Moreover, \(\Omega^\prime\) is sparse and block-diagonal, and applying the sparse Cholesky decomposition to each of its blocks allows us to invert it relatively easily.

Now consider concatenating the Walsh-Hadamard transform matrices, and their inverses, for each of the gates, which yields the block-diagonal Walsh-Hadamard transform matrix \(W\), and its inverse \(W^{-1}\).
As \(W^{-1}\) maps the gate eigenvalues to the gate probabilities, we use its inverse to calculate 
\begin{equation}\label{eq:probability-precision}
    \Sigma_{P}^{-1}=W\Sigma^{-1}W^{\intercal}.
\end{equation}

Our description of the transform matrices \(W\) and \(W^{-1}\) ignores an important subtlety.
The gate eigenvalue estimator vector we work with omits the identity eigenvalues for each gate.
Similarly, the gate probability estimator vector we project omits the identity probabilities for each gate.
The identity eigenvalues and probabilities ensure normalisation of the probability distributions for each gate.
We must omit these quantities to ensure that the corresponding covariance matrices have full rank.
However, \(W\) and \(W^{-1}\) act on versions of these vectors with those omitted elements and must be conjugated by appropriate transform matrices, which can be determined straightforwardly, to reflect these omissions.

At large scales, generalised least squares becomes impractical and we resort to weighted least squares, as we did here when performing ACES on the syndrome extraction circuit of a distance-\(25\) surface code.
In this case, we only generate the diagonal of \(\Omega^\prime\), which can easily be inverted.
Gate eigenvalue estimators are correlated within gate orbits because of the structure of the design matrix \(A\), whereas the off-diagonal elements of \(\Omega^\prime\) result from the simultaneous estimation of circuit eigenvalues.
Therefore, we expect that a version of \(\Sigma_P^{-1}\) calculated from only the diagonal of \(\Omega^\prime\) should suffice to ensure that Mahalanobis distance projection preserves the estimator covariance information necessary for relative precision estimation.
Indeed, we find this to be the case in practice.

We perform Mahalanobis distance projection into the probability simplex with the convex solver SCS~\citep{odonoghue_conic_2016, odonoghue_operator_2021}.
While this can be performed for the entire gate probability estimator vector, this quickly becomes practically intractable.
It suffices to perform the projection separately for each gate using its portion of the gate probability estimator vector and the corresponding block on the diagonal of \(\Sigma_P^{-1}\), which only has a minor impact on performance.

\end{document}